\documentstyle[buckow]{article}
\newcommand{\ky}{\kappa_{ \rm orb}}

\def\one{{\hbox{1\kern-.8mm l}}}

\def\beq{\begin{equation}}
\def\eeq{\end{equation}}

\begin {document}
\rightline{DSF- 1/2002}

\def\titleline{
Gauge Theories on Bound States of  Fractional Branes
\footnote{Talk given by the author at the RTN meeting ``The Quantum Structure 
of Spacetime and the Geometric Nature of Fundamental Interactions'',
 (Corfu`, September 2001)
and based on the papers in Ref.s ~\cite{d3d7, Bertolini:2000dk} done in collaboration with M. Bertolini, P. Di Vecchia, M. Frau, A. Lerda and I. Pesando.}}
\def\authors{ R. Marotta  }
\def\addresses{

Dipartimento di Scienze Fisiche, Universit{\`{a}} di Napoli
Complesso Universitario Monte S. Angelo, Via Cintia. I-80126 Napoli, Italy
\\
{\tt Raffaele.Marotta@na.infn.it} \\
}
\def\abstracttext{

We study the $N=2$ super Yang Mills theory living in the world volume  
of a bound state made of 
fractional D3/D7 branes at the orbifold ${\rm I\!R}^{1,5}\,\times\,
{\rm I\!R}^{4}/{\bf Z_2}$, by using the probe technique. 
We also discuss the boundary action for the system.}
\large
\makefront

\setcounter{footnote}{1}
\setcounter{page}{1}
\section{Introduction}
\label{1}

Since the formulation of the  Maldacena conjecture  
which states the equivalence between N=4 
super Yang-Mills theory in four dimensions and type IIB string theory 
compactified on $AdS_5 \times S_5$~\cite{mal} there have been many 
efforts to find new 
correspondences between string theories  and non-conformal and less 
supersymmetric gauge theories.
These attempts have  led to explore in detail and in  non trivial 
backgrounds the complementarity between the two-fold 
description of the low-energy properties of the D-branes,  one based on 
the 
classical curved  geometry generated by these non-perturbative extended 
objects  
and the other one given in terms of the
gauge theory living in their world-volume. 
According to this gauge/gravity correspondence, one can exploit the 
classical 
solution generated by  D-branes to study the gauge theory defined in 
their 
world-volume,
or, vice-versa to use the quantum properties of the gauge theory defined 
on 
D-branes to study their dynamics.
In particular in the last couple of years many efforts have been made to 
extend the Maldacena duality to non-conformal gauge theory with
$N~=~2$~\cite{Bertolini:2000dk, Petrini,Polchinski:2000mx,marco,GRANA2, dario, zaffa}
and $N=1$ supersymmetries~\cite{HKO, BDFM}. 

One of these approaches that is also the topic of this paper is based on 
the 
study of the gauge/gravity correspondence for a  bound state made of M D3  
 and N D7- fractional branes   living at the  singular point of 
the orbifold ${\rm I\!R}^{1,5}\,\times\,
{\rm I\!R}^{4}/{\bf Z_2}$\footnote{For a review on fractional 
branes see Ref.~\cite{BDM}}.
 
The gauge theory living  in the world-volume of this bound state is a 
non-conformal $N~=~2$ $SU(\mbox{M})$ super Yang-Mills with N hypermultiplets 
in the fundamental
representation of the gauge group and therefore such a bound state  represents 
one of  the 
simplest systems
to consider in order to extend the Maldacena conjecture.
Moreover  as shown in  Ref.s~\cite{d3d7}, the ``gravity dual" of this 
system possesses a
naked 
singularity of repulson type~\cite{kal}. This seems a general feature of all 
classical solutions 
corresponding to
non-conformal gauge theories and  according to the cosmic censorship 
conjecture they  must be rejected. However, in the case of the 
orbifold  there 
exists a general mechanism to resolve the singularity, in fact by probing
the 
background with a D3-fractional brane, one can see that the action 
of the probe becomes tensionless on a hypersurface 
called enhan\c{c}on that contains the naked singularity.
This aspect suggests that at the enhan\c{c}on locus new light degrees of 
freedom 
come into play and therefore the supergravity approximation  is not valid 
anymore; one also expects that string theory should resolve this problem.
According to this phenomenon  the region inside  the  enhan\c{c}on must be 
excised
obtaining a regular solution but in a region of the space that is far 
from the brane. 
Therefore for these kinds of solutions, it is
not
simple to take  the decoupling limit  and studying  physics inside the 
enhan\c{c}on is still an interesting and opened question ( for a discussion on the 
physics inside the enhan\c{c}on see Ref.s~\cite{Aharony, Petrini:2001fk}.

\section{D3/D7 bound state and its classical solution}
\label{2}

In the first part of this section we will analyze in detail 
the D7 branes of type 
IIB at the orbifold ${\rm I\!R}^{1,5}\,\times\,
{\rm I\!R}^{4}/{\bf Z_2}$. We closely follow the approach discussed in Ref. 
\cite{d3d7}  and in particular we discuss 
the case of a  D7 brane completely extended along the orbifold 
since this is the 
relevant configuration to yield four-dimensional N=2
gauge theories in the presence of fractional D3 branes.
The peculiar aspect of the D7 branes extended along the orbifold
is that they are sources not only of 
those ``untwisted'' fields typical of a D7 brane, but also  of
``twisted'' scalar fields $b$ and $c$ defined as follows: 
\begin{equation}
C_{2} = c~ \omega_2~~~~~,~~~~B_{2} = b~ \omega_2
\label{wra98}
\end{equation}
where $B_2$ and $C_2$ are respectively the  Kalb-Ramond and  the 
RR  two form potentials, while
$\omega_2$ is the anti-self dual 2-form
associated to the vanishing 2-cycle of the orbifold ALE space.

It is interesting to observe that these twisted fields are 
the same of those emitted by the fractional D3 branes studied 
for examples in Ref.s~\cite{Bertolini:2000dk,Polchinski:2000mx}.

These features can be clearly seen by computing the vacuum 
energy $Z$ of the open string stretched between two such D7
branes that is given by:
\begin{equation}
Z = \int_0^\infty \frac{d s}{ s} ~{\rm Tr}_{\rm NS-R} \left[P_{\rm GSO}
\left(\frac{\one + g}{2} \right)  {\rm e}^{- 2 \pi s (L_0-a)} \right]
\label{part45}
\end{equation}
where $P_{\rm GSO}$ is the GSO projection, $g$ is the orbifold ${\bf Z}_2$
parity, and $a=1/2$ in the NS sector and $a=0$ in
the R sector.
When one takes the $\one$ inside the bracket, one gets half of
the usual contribution of the open strings stretched between two D7 branes 
in flat space, whereas as  one
takes the $g$ inside the bracket one obtains 1/16 times the contribution
of the twisted sectors of the fractional D3 branes of 
Ref.s~\cite{Bertolini:2000dk,Polchinski:2000mx}
After performing the modular transformation $s\to 1/s$ 
and factorizing the resulting expression in the closed string channel,
one can derive the boundary state associated to the D7
brane along the orbifold. The final result is  the sum of two terms:
\begin{equation}
|{\rm D}7\rangle = |{\rm D}7\rangle^{\rm U}+|{\rm D}7\rangle^{\rm T}
\label{bsd7}
\end{equation}

The untwisted part $|{\rm D}7\rangle^{\rm U}$ is the same as that of the 
D7 branes in flat space but with a normalization differing by a 
factor of $1/\sqrt{2}\,$; the twisted part $|{\rm D}7\rangle^{\rm T}$ is, 
instead, similar to that of the fractional D3 branes 
but with a normalization differing
by a factor of 1/4. 
By saturating the boundary state (\ref{bsd7})
with the massless closed string states of the various sectors, one can 
determine which are the
fields that couple to the fractional D7 brane.

{F}rom the explicit couplings it is
possible to infer the form of the terms of the world-volume
action that are linear in 
the bulk fields obtaining the following expression:

\begin{eqnarray}
S_{~\rm b}^{\rm D7} &\simeq& 
-\,\tau_7
\, \int d^8x~{\rm e}^\phi\,\sqrt{-\det g_{ab}} ~+~
\tau_7\, \int
C_{8}
\nonumber\\
&+& \frac{\tau_3}{2(2\pi\sqrt{\alpha^{'}})^2}
\,\int d^4x~\sqrt{-\det
g_{\alpha\beta}}~{\widetilde b}\,-\,
\frac{\tau_3}{2(2\pi\sqrt{\alpha^{'}})^2}\,\int A_4+...
\label{sbtw}
\end{eqnarray}

where $\tau_p\equiv T_p/({\sqrt{2}\,\ky})$ and $g_{ab}$ is the induced 
metric. The previous expression represents the whole information that 
can be extracted from the boundary state but it is enough to compute
the classical solution generated by a bound state made of D3/D7 fractional 
branes.

In order to determine the 
complete boundary action for our D7 brane completely extended along the 
orbifold we have first to compute the classical solution and after, by 
imposing the zero force condition for the system, we will be able to 
fix the higher order couplings appearing in the boundary action.
However this is exactly the argument of the forthcoming discussion.

In the last part of this section we compute the classical solution
of the type IIB supergravity equation of motion, 
for a bound state made  of M 
D3-fractional 
branes and N D7-branes on the orbifold ${\rm I\!R}^{1,5}\,\times\,
{\rm I\!R}^{4}/{\bf Z_2}$.
In particular we will consider configurations in which the D7 branes are 
entirely   extended along the orbifold; i.e. in the directions 
 $x^0\dots x^3$ and in the orbifolded directions $x^6\dots x^9$, while 
the D3-branes are transverse to the 
orbifold and therefore extended along the directions $x^0\dots x^3$.

The starting point are the equations of  motion of type IIB supergravity 
in ten dimensions that, for the fields that we are interested in, can be 
combined in two complex and compact expressions given by:

\begin{equation}
d {}^{*}d \tau + i {\rm e}^{\phi} d \tau \wedge {}^{*} d \tau +
\frac{i}{2} G_3 \wedge {}^{*}  G_3 = 2i \kappa^2 {\rm e}^{- \phi}
\left[ \frac{\delta {\cal{L}}_b}{\delta \phi} +i {\rm e}^{-\phi}
\frac{\delta {\cal{L}}_b}{\delta C_0} \right]
\label{dileq45}
\end{equation}
and
\begin{equation}
d {}^{*} G_3 + d \tau \wedge \left[i {\rm e}^{\phi} {}^{*} G_3 +
  {}^{*} H_3  \right] - i {\tilde{F}}_5 \wedge G_3 = - 2i \kappa^2
  \left[ \frac{\delta {\cal{L}}_b}{\delta B_2} - \tau \frac{\delta
  {\cal{L}}_b}{\delta C_2} \right]
\label{g3eq67}
\end{equation}

where $G_3= dC_2+\tau dB_2$, $\tau= C_0 + i \mbox{e}^{-\phi}$ is 
the standard combination of the axion and dilaton fields and 
${\cal{L}}_b$ is the boundary lagrangian for our bound system.
The equation of motion for the other fields that couple to 
our system are not relevant for the aim of this paper  and therefore they 
will not be taken into account in the following discussion.
The solution of the previous system of equations of motion 
compatible with the symmetries of the system and preserving  
eight real supersymmetries is:

\begin{equation}
\tau =  {\rm i}\left(1 -\, \frac{Ng_s}{2\pi}\, \log \frac{z}{\epsilon}\right)
~~, 
\label{tausol}
\end{equation}
and
\begin{equation}
\gamma = \,{\rm i}\,2\pi\alpha' g_s\,\left[\frac{\pi}{g_s} + 
(2M-N)\,\log \frac{z}{\epsilon}\right]
\label{gammasol}
\end{equation}
where $z=x^4\, +\, i\, x^5$ and  we have chosen the integration constants to enforce the
appropriate background values.
It is interesting to observe that, by rewriting the solution in terms of the 
real component fields, and expanding them 
in powers of $\lambda= N\,g_s$ we realize that  $C_0$ does not receive 
corrections to higher orders in $\lambda$ while the 
twisted fields $b$ and $c$ acquire an infinite tail of logarithmic 
terms. This is to be contrasted with the solution of the
pure fractional D3 branes \cite{Bertolini:2000dk,Polchinski:2000mx}
where the twisted scalars had, instead, only terms at first order.
Thus, if one 
wants to determine
the classical profile of the twisted scalars using the boundary state 
formalism in the presence of fractional D7 branes, 
it is not sufficient to consider contributions with just 
one boundary, but it is necessary to sum over 
all contributions with an arbitrary number of boundaries as explained 
in Ref.~\cite{nbps}, which, 
due to the open/closed string duality, is equivalent 
to sum over an arbitrary number of open-string loops.  

Having the complete solution, we can verify the no-force condition
and check the structure of the world-volume action of the bound state. 
If we substitute our solution in the boundary action of the 
D7-brane given in the eq.(\ref{sbtw}), we see the missing of zero force 
condition.
This problem is overcome by remembering that the boundary action   
(\ref{sbtw}) only contains the linear coupling emerging from the boundary 
states. By imposing the zero force condition and remembering that
the D7 branes being extended along the orbifolded direction  couple also 
with 
the twisted fields, it is  possible to deduce all the higher order couplings,
obtaining:

\begin{eqnarray}
S_{~\rm b}^{\rm D7}&=&-\,\tau_7
\, \int d^8x~{\rm e}^\phi\,\sqrt{-\det g_{ab}} ~+~
\tau_7\, \int
C_{8}\nonumber\\
&+&\,\frac{\tau_3}{2(2\pi\sqrt{\alpha^{'}})^2}
\int d^4x~\sqrt{-\det
g_{\alpha\beta}}~{\widetilde b}
\left(1+\frac{\widetilde b}{4\pi^2\alpha'}\right)-
\frac{\tau_3}{2(2\pi\sqrt{\alpha^{'}})^2}
\int A_4
\,\left(1+\frac{\widetilde b}{4\pi^2\alpha'}\right)\nonumber \\
&-&
\frac{\tau_3}{2(2\pi\sqrt{\alpha^{'}})^2}
\int C_4\,\,
{\widetilde b}\,\left(1+\frac{\widetilde b}{4\pi^2\alpha'}\right)~~.
\label{sbtwfin}
\end{eqnarray}

It would be also interesting to confirm the 
boundary action  by string calculations  or by geometrical
considerations.

\vskip 1cm
\section{The probe action and the ${\cal N}=2$ gauge theory}
\label{section3}
 
The supergravity solution found in the previous section can
provide, through the probe technique, non-trivial information on its dual 
four-dimensional gauge theory. 

According to  the probe technique,  we consider a fractional D3 brane
 carrying a gauge field $F_{\mu\nu}$ slowly moving in a given supergravity 
background. By studying the  boundary action of this probe we
can get  non trivial information on  the 
Coulomb phase of the $SU(M+1)$ gauge theory 
 broken to $SU(M)\times U(1)$ that corresponds to taking 
the probe at a distance $\rho=|z|$ from the source. 

Applying this technique  to the background 
 produced by our  bound state made of 
 M D3 and N D7 fractional branes, we find that in the probe action 
all  the dependence on the untwisted fields 
 drops out only leaving the following expression:
 
 \begin{equation}
 S_{gauge} = -\,\frac{1}{g_{\rm YM}^{2} (\mu)}
 \int d^4 x  \left\{ \frac{1}{2} \partial_{\alpha} \Phi^i
 \partial^{\alpha} \Phi^i + \frac{1}{4} F_{\alpha \beta} F^{\alpha \beta}
   \right\} 
  +  \frac{\theta_{\rm YM}}{32 \pi^2} \int d^4 x F_{\alpha \beta} 
  {\tilde{F}}^{\alpha \beta} 
  \label{bound53}
  \end{equation} 
  where
  \begin{eqnarray}
  \frac{1}{g_{\rm YM}^{2} (\mu) } &=& \frac{1}{g^2_{\rm YM}} + 
  \frac{2M-N}{8 \pi^2} \log \mu  \hspace{.5cm}; \hspace{.5cm}
  g^{2}_{\rm YM} = 8\pi g_s    
  \label{runn23} \\
  \nonumber \\
  \theta_{\rm YM} &=& (2M-N) \; \theta 
  \label{runn24}
  \end{eqnarray}
are the effective Yang-Mills gauge coupling and $\theta$-angle, respectively,
while $\Phi^i= (2\pi\alpha^{'})^{-1}\, x^i$ being $x^i$ the two coordinates 
transverse both to the probe and to the orbifold. 

It is interesting to observe that the coefficients in front the two 
kinetic terms in the expression (\ref{bound53})  are the same, this is in 
agreement  
with the fact that the gauge theory living on the brane has  
 ${\cal{N}}=2$ supersymmetry.
 
Eq.(\ref{runn23}) clearly shows that 
$g_{\rm YM} (\mu)$ is the running coupling constant of
an ${\cal{N}}=2$ supersymmetric gauge theory 
with gauge group $SU(M)$ and $N$ hypermultiplets in the fundamental 
representation, while the  renormalization group scale is related to the 
supergravity variables by: $\mu\equiv |z|/\epsilon$. Indeed this is the field 
theory 
 living on the system of $M$ D3-branes and $N$ D7-branes.

 Furthermore by studying the boundary action of a fractional D3-probe
in the background defined by the eq.s ~(\ref{tausol}) and (\ref{gammasol})
 one sees that on the geometric locus:
 \begin{equation}
 |z_e| \;=\; \epsilon \;{\rm e}^{- \pi\,/\,(2M-N)\,g_s}~~,
 \label{enh}
 \end{equation}
 the probe becomes tensionless, thus indicating the presence of 
 an enhan\c{c}on. At distances smaller 
 than $|z_e|$ the probe has negative tension, while at the 
 enhan\c{c}on extra light degrees 
 of freedom come into play~\cite{Polchinski:2000mx}. This means 
 that the supergravity approximation leading to the solution described 
 in section 2 is not valid anymore, and that the region 
 of space-time $\rho < |z_e|$ is excised. 

A distinctive feature of the D3/D7 system 
with respect to that of pure fractional 
D3-branes of Ref.~\cite{Bertolini:2000dk} is 
that the twisted scalars given in the eq.s (\ref{tausol}) and (\ref{gammasol})
are expressed as an infinite series in the open string coupling. 
However, the scalar field combinations which have 
a meaning at the gauge theory level are again exact at one-loop, as 
expected for a ${\cal N}=2$ super Yang-Mills theory. This non-trivial 
cancellation is a (higher loop) check of the validity 
of the gauge/gravity correspondence. 

In conclusion we have seen that by using the classical solution generated by 
bound 
states of fractional branes it is possible to have information on the 
perturbative region of the moduli space of $N=2$ supersymmetric gauge theory.

These solutions, also for the presence in the infrared of a naked singularity, 
prevent
the possibility to perform the decoupling limit, and therefore 
they seem in contrast 
with a duality interpretation \'a la Maldacena where the supergravity 
solution gives a good description of the gauge theory at large 't Hoof 
coupling. 
Our results instead can be easily  understood if we regard the classical 
supergravity 
solution as a large distance expansion around the asymptotic flat background.
The curvature is clearly small 
and the supergravity is reliable, while the expansion, as explained in
Ref.~\cite{nbps}, can be seen as a sum over the tree level close string 
diagrams, or because the open/closed string duality, as a sum over 
open string loops.
From this point of view,  the supergravity solutions  are  
considered as an   
effective way of summing over open string loops 
and therefore they can only encode the 
perturbative properties of the ${\cal{N}}=2$
gauge theory living on the world-volume of a fractional D3-brane.



\begin{thebibliography}{99}
 \bibitem{d3d7} 
 M.~Bertolini, P.~Di~Vecchia, M.~Frau, A.~Lerda and R.~Marotta, 
 \emph{${\cal{N}}=2$ gauge theories on systems of fractional D3/D7 branes},
 Nucl. Phys. {\bf B261} (2002) 157-178, {\tt hep-th/0107057}.
\bibitem{Bertolini:2000dk}
M.~Bertolini, P.~Di~Vecchia, M.~Frau, A.~Lerda, R.~Marotta and
I.~Pesando, \emph{Fractional D-branes and their gauge duals},
JHEP {\bf 02} (2001) 014, {\tt hep-th/0011077}.

\bibitem{mal}
 J. Maldacena, \emph{The large N limit of superconformal field
 theories and supergravity}, Adv. Theor. Math. Phys. {\bf 2} (1998)
 231, {\tt hep-th/9711200}.
 %
 \bibitem{Petrini} 
  N. Evans, C.V. Johnson and M. Petrini, \emph{The Enhancon and N=2 Gauge Theory/Gravity RG Flows}, 
 JHEP {\bf 10} (2000) 022, {\tt hep-th/0008081}. 
 %
 \cite{Polchinski:2000mx}
 \bibitem{Polchinski:2000mx} J.~Polchinski, \emph{N = 2 gauge-gravity
 duals}, Int. J. Mod. Phys.  {\bf A16} (2001) 707, {\tt hep-th/0011193}.
 %
 \bibitem{marco} M. Bill\`o, L. Gallot and A. Liccardo, {\em Classical
   geometry and gauge 
  duals for fractional branes on ALE spaces}, {\tt hep-th/0105258}.
 %
 \bibitem{GRANA2}
 M. Grana and J. Polchinski,
 {\em Gauge/gravity duals with holomorphic dilaton}, 
 {\tt hep-th/0106014}.
 %
 \bibitem{dario} J.P. Gauntlett, N. Kim, D. Martelli, D. Waldram, 
 \emph{Wrapped fivebranes and N=2 super 
 Yang--Mills theory}, {\tt hep-th/0106117}.
 %
 \bibitem{zaffa} F. Bigazzi, A.L. Cotrone, A. Zaffaroni, 
 \emph{N=2 gauge theories from 
  wrapped five-branes}, {\tt hep-th/0106160}.
 %
 \bibitem{HKO}
 C. P. Herzog, I. R. Klebanov, P. Ouyang, \emph{ Remarks on the Warped 
 Deformed Conifold}, I.R.K.'s talks at the Lisbon School on Superstrings 
 II, July 2001 and at the Benasque Workshop, July 2001, {\tt hep-th/0108101}.
 %
 \bibitem{BDFM} M. Bertolini, P. Di Vecchia, G. Ferretti and R. Marotta,
 \emph{Fractional Branes and $N=1$ Gauge Theories}, {\tt hep-th/0112187}. 
 %
\bibitem{BDM} M. Bertolini, P. Di Vecchia and R. Marotta,
\emph{$N=2$ For-Dimensional Gauge Theories from Fractional Branes},
{\tt hep-th/0112195}.
%
 \bibitem{kal} R. Kallosh and A. Linde, 
 {\em Exact supersymmetric massive and massless white holes},
 Phys. Rev. {\bf D52} (1995) 7137, {\tt hep-th/9507022}.
 \bibitem{Aharony}
  O.~Aharony, \emph{A note on the holographic interpretation of
  string theory backgrounds with varying flux}, JHEP {\bf 03}
  (2001) 012, {\tt hep-th/0101013}.
  %
  \bibitem{Petrini:2001fk}
  M.~Petrini, R.~Russo and A.~Zaffaroni, \emph{N = 2 gauge theories
  and systems with fractional branes}, {\tt hep-th/0104026}.
 %
 \bibitem{nbps}
 M. Bertolini, P. Di Vecchia, M. Frau, A. Lerda, R. Marotta, R. Russo
 \emph{Is a classical description of stable non-BPS D-branes possible?}, 
 Nucl. Phys. {\bf B590} (2000) 471, {\tt hep-th/0007097}.
 %

\end{thebibliography}
\end{document}